# Interlayer Quasi-Bonding Interactions in 2D Layered Materials: A Classification According to the Occupancy of Involved Energy Bands

Yuan-Tao Chen,[1,2,†] Peng-Lai Gong,[1,2,†] Yin-Ti Ren,[2] Liang Hu,[2] Hu Zhang,[1] Jiang-Long Wang,[1] Li Huang,[2,3,*] and Xing-Qiang Shi[1,*]

[1] Key Laboratory of Optic-Electronic Information and Materials of Hebei Province, Institute of Life Science and Green Development, College of Physics Science and Technology, Hebei University, Baoding 071002, P. R. China

[2] Department of Physics, Southern University of Science and Technology, Shenzhen 518055, P. R. China

[3] Guangdong Provincial Key Laboratory of Energy Materials for Electric Power, Shenzhen, 518055, P. R. China

[†]These authors contributed equally to this work
[*]E-mails: shixq20hbu@hbu.edu.cn, huangl@sustech.edu.cn

**Abstract:** Recent studies have revealed that the interlayer interaction in two-dimensional (2D) layered materials is not simply of van der Waals character but could coexist with quasi-bonding character. Here we classify the interlayer quasi-bonding interactions into two main categories (I: homo-occupancy interaction, II: hetero-occupancy interaction) according to the occupancy of the involved energy bands near the Fermi level. Then we investigate the quasi-bonding-interaction-induced band structure evolution of several representative 2D materials based on density functional theory calculations. Further calculations confirm that this classification is applicable to generic 2D layered materials and provides a unified understanding of the total strength of interlayer interaction, which is a synergetic effect of the van der Waals attraction and the quasi-bonding interaction. The latter is stabilizing in main category II and destabilizing in main category I. Thus, the total interlayer interaction strength is relatively stronger in category II and weaker in category I.

TOC graphic entry

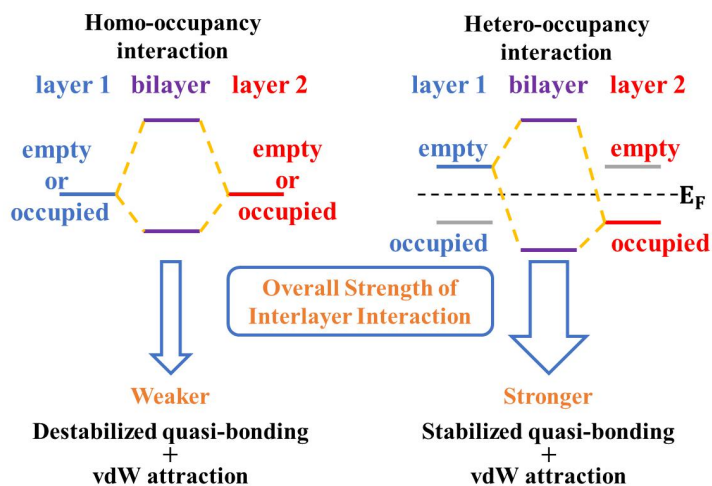

Since the exfoliation of graphene from graphite,[1] two-dimensional (2D) layered materials, such as transition metal dichalcogenides (TMDs),[2-4] black phosphorus,[5] hexagonal boron nitride (*h*-BN),[6-7] the InSe family,[8-10] the $Bi_2O_2Se$ family[11] and recently the $MoSi_2N_4$ family,[12-13] have attracted increasing attention for their extraordinary properties and potential applications.[14] Many recent studies focus on 2D materials' van der Waals (vdW) homo- and hetero-structures. It has been demonstrated that the interlayer coupling is an essential degree of freedom to manipulate the electronic,[15-18] thermoelectric[19-21] and magnetic properties,[22] and promote the emerging field of interlayer engineering of 2D layered materials.[9, 23-26]

Few-layer 2D materials show layer-number-dependent properties.[27-30] As a pioneering work, Zhao *et al.* reported the relatively strong interlayer interaction in 2D layered $PtS_2$, which is much stronger than the typical vdW interaction.[31] The interlayer interaction of $PtS_2$ is named "covalent-like quasi-bonding" for its covalent-bond-like charge accumulation between the S atoms of adjacent $PtS_2$ layers, although the magnitude of the accumulated charge is much smaller than that in the typical chemical bonding.[31] The charge accumulation in the interlayer region was confirmed in the synchrotron X-ray diffraction experiment for a similar system, $TiS_2$.[32] This kind of interlayer quasi-bonding interaction plays an essential role in the band structure evolution of few-layer $PtS_2$ as the layer-number varies. The typical interlayer interaction (with a total strength of ~20 meV/Å$^2$) is also crucial in tuning 2D materials' properties,[33] partially due to the coexistence of interlayer quasi-bonding interaction. As well-known examples, $MoS_2$ experiences a direct-indirect band gap transition from monolayer to bilayer,[34-35] black phosphorus has a tunable band gap of 1.51 eV to 0.36 eV from monolayer to bulk.[36] These interlayer-interaction-induced band evolution arise from, in addition to the quantum confinement effect,[37] the interlayer quasi-bonding, which leads to the down-shifts of conduction band minimum (CBM) and/or the up-shifts of valence band maximum (VBM) which are the result of interlayer interaction.[31, 38-40] Recent study reveal that the interlayer interaction could cause the unusual up-shift of CBM and down-shift of VBM and thus increases band gap.[41] As examples discussed below, 2D triphosphides, with a chemical formula of $XP_3$ (X is a group III-A, IV-A or V-A element), have gained increasing attention in recent years.[20, 37, 42-46] In these studies of few-layer triphosphides and their heterostructures,[16-18] the interlayer interaction has been shown to be stronger, and they can induce metal-semiconductor transitions with increasing thickness.[17]

Here, we extend the frontier orbitals perspective (as shown in Figure S1)[47] that is used to analyze the interatomic or intermolecular interaction to deal with the interlayer quasi-bonding interaction of 2D layered materials. Then we classify the interlayer interactions of 2D materials into two main categories.

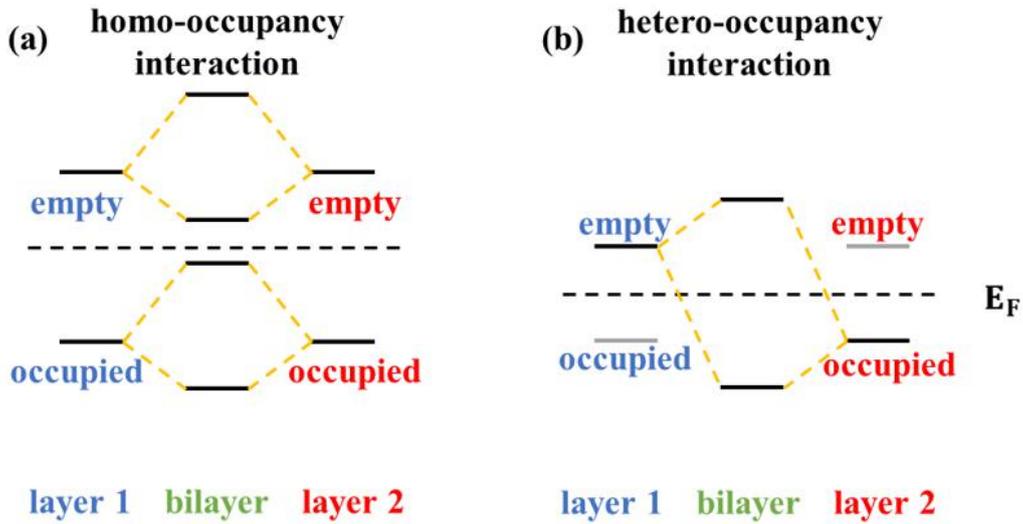

**Figure 1. Classification of interlayer interactions.** (a) homo-occupancy interaction with subtypes of empty-empty and occupied-occupied interactions; (b) hetero-occupancy interaction with an example of empty-occupied interaction. Note that the subtypes displayed here are not complete, and partially occupied bands could also occur.

Category I: The homo-occupancy interaction arises from the interaction of orbitals with the same occupancy, for example, the hybridization between two empty or two occupied orbitals (empty-empty or occupied-occupied interactions). As Figure 1a shows, the anti-bonding state of occupied-occupied interaction may push the VBM of the bilayer to higher energy, while the bonding state of empty-empty interaction could push the CBM lower, leading to a band gap decrease in bilayer. Usually, the up-shift of VBM is larger than the down-shift of CBM.[31, 39-40] Also, considering that the up-shift of the anti-bonding state is larger than the down-shift of the bonding state, the occupied-occupied interaction costs energy. Thus, this kind of interlayer interaction shows relatively small binding energy, as a competition of vdW attraction between layers (that lowers the total energy) and interlayer quasi-bonding 'repulsive' interaction from the occupied-occupied interaction (that costs energy). In addition, there may be a semiconductor-metal transition if the up-shifted VBM from occupied-occupied interaction overcomes the down-shifted CBM from empty-empty interaction.

Category II: As Figure 1b shows, the hetero-occupancy interaction between two orbitals with different occupancies, such as the empty-occupied interaction. The empty-occupied interaction does not induce semiconductor-metal transition. In the case of the CBM-VBM interaction (a typical empty-occupied interaction), it can enlarge the band gap.[41] As exceptional cases, interlayer interactions with unsaturated orbitals (partially occupied bands) possibly lead to semiconductor-metal transition, which will be discussed later.

To demonstrate the conceptual picture of the interlayer interactions, we choose arsenene,[48-54] $SnP_3$ [55-56] and $InP_3$ [45, 57-58] as representative 2D few-layer materials, which have different numbers of valence electrons filling the energy bands around Fermi energy. Due to the decreasing occupancy of the involved bands in these materials (phosphorus and arsenic belong to group V-A, tin belongs to group IV-A, indium belongs to group III-A), the interaction types change in these three materials. Specifically, as shown in Figure 2, from electron counting, each X atom in $XP_3$ (or As atom for arsenene) forms three chemical bonds in the intralayer, leaving two (As, P), one (Sn) and zero (In) electrons along the out-of-plane direction, so that As and P may contribute fully-occupied orbitals, Sn may contribute partially-occupied orbitals, and In may contribute empty orbitals. As a result, there are different types of X--P (or As--As) interlayer interactions in these three systems. As we will show later, the interlayer interaction in $InP_3$ not only have the interlayer In--P interaction but also the interlayer In--In interaction, suggesting that subtypes of interlayer interaction may coexist. The details of interlayer interaction in $InP_3$ are displayed in Figure S3. The optimized monolayers of the three materials all exhibit buckling hexagonal honeycomb configurations, which ensures the environment of interlayer interaction is similar. Thus, the occupancy of the involved bands dominates in the interlayer quasi-bonding interaction. In both triphosphides, the metal atoms (In and Sn) are distorted in the out-of-plane direction relative to the phosphorus atoms but behave differently. Specifically, tin (Sn) atoms in relaxed monolayer $SnP_3$ are distorted to the outside of the layer. In contrast, the indium (In) atoms in the relaxed monolayer $InP_3$ are distorted to the inside of the layer relative to the phosphorus atoms.

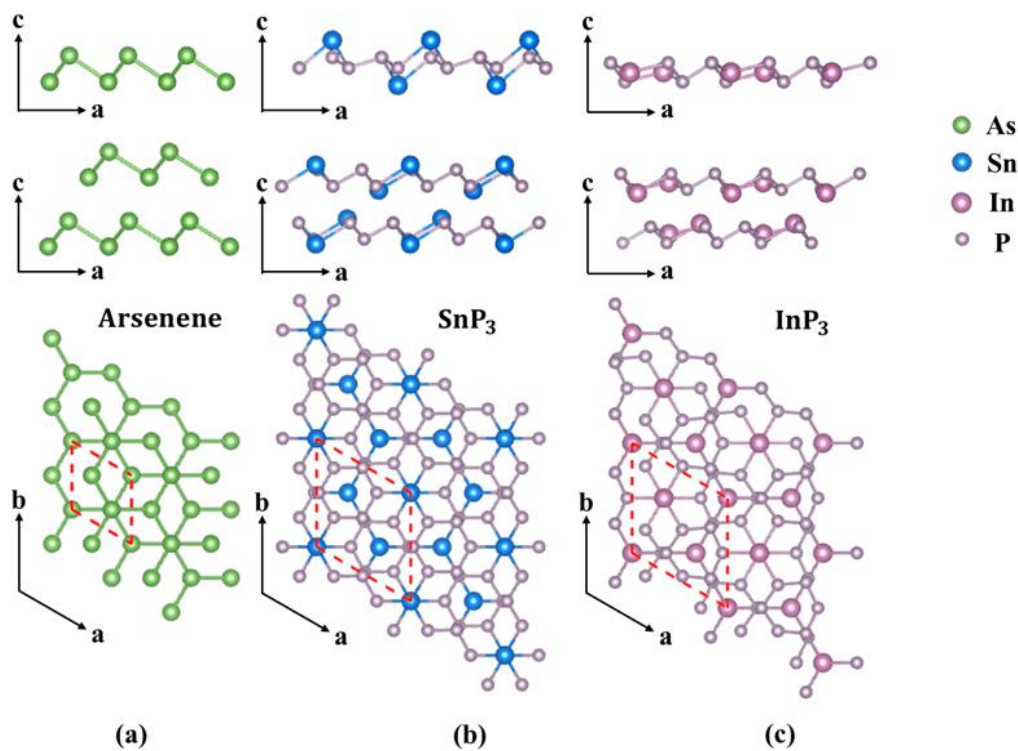

**Figure 2. Interlayer interaction induced geometric structure change from relaxed monolayer to bilayer.** Side-views of relaxed monolayer (upper panels), side- and top-views of bilayer (middle & bottom panels). (a) arsenene, (b) $SnP_3$ and (c) $InP_3$. The rhombus dashed lines in the top views represents the primitive cell.

The side- and top-views of bilayer arsenene, $SnP_3$ and $InP_3$ are shown in the middle and bottom panels of Figure 2. At the interlayer region of bilayer structures, the distortions of metal atoms relative to phosphorus in $SnP_3$ and $InP_3$ are weakened relative to the relaxed monolayer structure. More details of the geometric structures including lattice constants can be found in Supporting Information. The significant interlayer-interaction-induced geometric changes indicate that the interlayer interactions in triphosphides are much stronger than that in bilayer arsenene. To exclude the impact of geometric changes, we take the unrelaxed single-layer structure in the bilayer as an intermediate state to probe the band structure evolution from fully optimized monolayer to bilayer. To avoid confusion, we use *unrelaxed single-layer* to describe the single-layer structure directly from bilayer and *relaxed monolayer* for the optimized single-layer structure.

*Homo-occupancy interlayer interaction in arsenene.* As shown in Figure 2a, the geometric structure change of arsenene is negligible between the relaxed monolayer and the unrelaxed single-layer structure. The band structures of unrelaxed single-layer and bilayer arsenene are shown in Figure 3, the projected crystal orbital Hamiltonian population (pCOHP) is also presented.

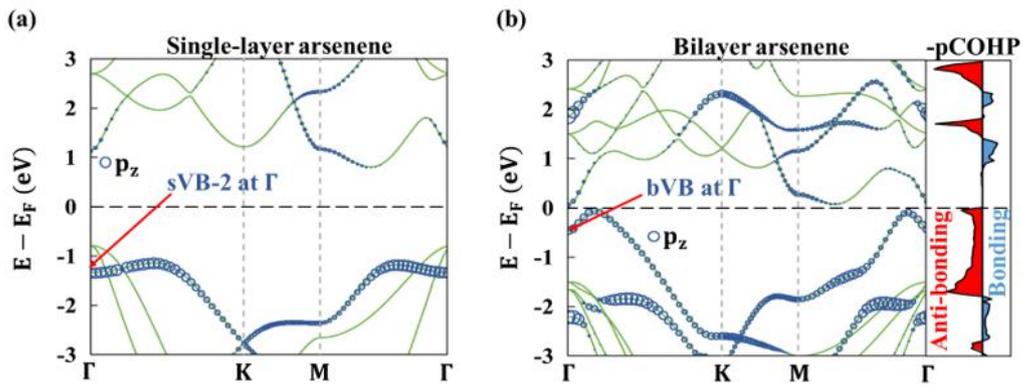

**Figure 3. Interlayer interaction induced band gap decrease in arsenene.** (a) Projected band structure of unrelaxed single-layer arsenene. (b) The band structure of bilayer arsenene (left panel) and the pCOHP of interlayer As--As interactions (right panel). The "s" and "b" represent unrelaxed single-layer and bilayer, respectively.

As Figure 3 shows, in unrelaxed single-layer arsenene, the $p_z$ orbitals of arsenic make the main contribution to the sVB-2 at Γ, and the $p_z$ orbitals of arsenic are the major participators of interlayer interaction. In Figure 3b, the anti-bonding state of interlayer As--As interaction becomes bVB at Γ. It is consistent with our discussion above about homo-occupancy interaction and matches the picture of the band evolution in other 2D materials with the homo-occupancy (category I) interactions.[31, 38-40] It is worth mentioning that the interlayer interaction leads to a semiconductor-metal transition from bilayer to trilayer arsenene. More details can be found in Figure S6.

*Hetero-occupancy interlayer interaction in* $SnP_3$. As mentioned above, there is a significant structural change from relaxed monolayer to bilayer $SnP_3$. Interestingly, as shown in Supporting Information Section VIII, the relaxed monolayer is semiconducting with a band gap of around 0.42 eV, while the unrelaxed single-layer structure is metallic. As an element of group V-A (group IV-A), each P (Sn) atom in unrelaxed single-layer $SnP_3$ forms three intralayer bonds, leaving two (one) electrons forming lone pairs (unsaturated orbital) in the out-of-plane direction.[59] Thus, the major interlayer interaction in $SnP_3$ is a special case of the hetero-occupancy interaction that involves partially-occupied bands.

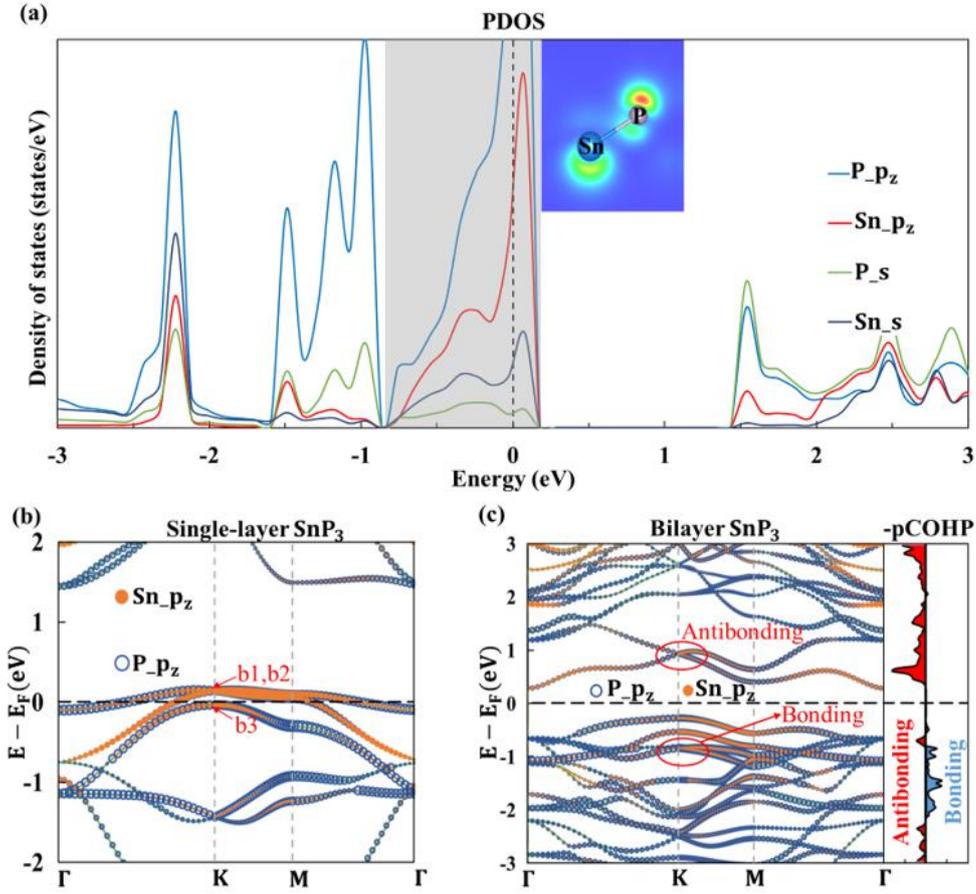

**Figure 4. Interlayer-interaction-induced band evolution in $SnP_3$.** (a) PDOS of the unrelaxed single-layer $SnP_3$, the insert shows the partial charge density in the shaded energy window around Fermi level from $-0.83$ eV to 0.18 eV. (b) The projected band structure of unrelaxed single-layer $SnP_3$. The energy bands involved in interlayer interaction from unrelaxed single-layer to bilayer are noted as b1, b2 and b3. (c) Projected band structure and the pCOHP of interlayer Sn--P interaction in bilayer $SnP_3$. The bonding states and anti-bonding states of Sn--P interaction are marked by red ellipses. The Fermi energies are all set to zero.

Figure 4a shows the projected density of states (PDOS) of the unrelaxed single-layer $SnP_3$. The partial charge density (shown in the insert) in the shaded energy window around the Fermi level (from $-0.83$ eV to 0.18 eV) indicates that the out-of-plane charge distribution of Sn and P make the main contribution to the bands around the Fermi level, which are denoted as b1, b2 and b3 in Figure 4b. Specifically, considering the unsaturated occupation of Sn atoms and the full occupation of P atoms, the partially occupied bands b1 and b2 are mainly contributed by the unsaturated out-of-plane orbitals of Sn atoms and the fully occupied band b3 is mainly contributed by P atoms.

In the formation of bilayer $SnP_3$, the out-of-plane orbital-interaction of tin and phosphorus atoms is the primary origin of the interlayer interaction. Thus, in bilayer $SnP_3$, the occupied orbitals of P (the band b3 in Figure 4b) and the unsaturated orbitals of Sn (the bands b1 & b2 in Figure 4b) form bonding and anti-bonding states due to the hetero-occupancy interaction. Thus, one of the partially-occupied band becomes empty, and the other partially-occupied band becomes fully occupied. More details can be found in Supporting Information. Since there are two groups of Sn--P interactions in the interlayer region per unit cell of bilayer $SnP_3$ (one Sn from each layer, refer to the geometric structures in Figure 2b), there are two bonding and anti-bonding states in the band structure of bilayer $SnP_3$ (Figure 4c), which are originated from the interlayer Sn--P interactions. Considering the occupancy change around Fermi energy, the hetero-occupancy interactions with unsaturated bands possibly cause a metal-semiconductor transition.

*Combination of homo- and hetero-occupancy interactions in $InP_3$.* As shown in Figure 2c, the indium atoms in the relaxed monolayer $InP_3$ are distributed in the middle of the layer, and their distortions relative to the vertical position of phosphorus are weakened in bilayer. It should be noted that in bilayer $InP_3$, the indium atoms at the interlayer region (which mainly contribute to the interlayer interaction) are different from the indium atoms away from the interlayer region. In order to distinguish between these two types of indium atoms, we use 'i' to indicate the indium atoms at the interlayer region, and o' is used for another type of indium atoms. These notations are also used for the corresponding atoms in unrelaxed single-layer $InP_3$.

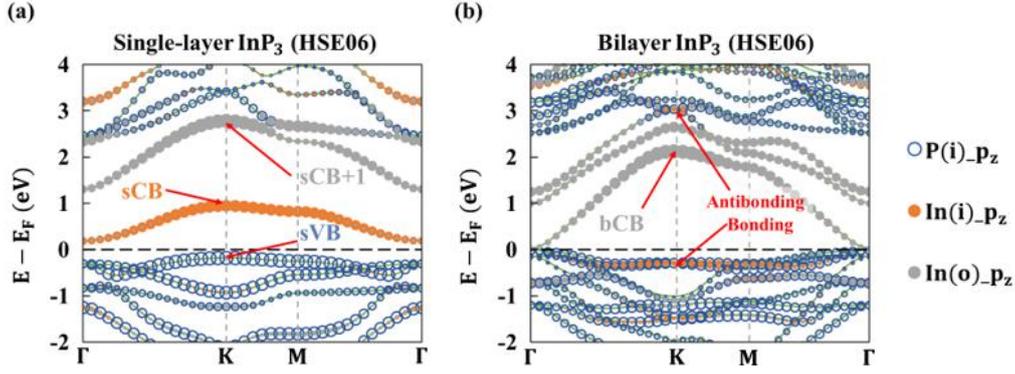

**Figure 5. Band structure evolution from unrelaxed single-layer to bilayer $InP_3$.** Projected band structures calculated under HSE06 of (a) unrelaxed single-layer and (b) bilayer $InP_3$. In (b), the energy bands of the bonding (bVB &bVB-1) and anti-bonding (bCB+2 & bCB+3) states of the interlayer In--P interactions are denoted. The 's' and 'b' represent the unrelaxed single-layer and bilayer, respectively. The 'i' represents the indium atoms in the interlayer region and 'o' represents the other indium atoms.

Roughly speaking, in the formation of bilayer $InP_3$, the $p_z$ orbitals of In(i) (sCB in Figure 5a, empty) and P (sVB in Figure 5a, occupied) hybridize and form bonding and anti-bonding states which are denoted in Figure 5b. According to our discussion and previous research,[41] this hetero-occupancy interaction (empty-occupied interaction) possibly leads to a band gap increase when the involved energy bands are VBM and CBM. However, there is band gap decrease from unrelaxed single-layer to bilayer $InP_3$. It is the result of the interlayer In--In interaction. The $p_z$ orbitals of In(o) (both empty, sCB+1 in Figure 5a) hybridize and the bonding state lower the bCB (Figure 5b). Details could be found at Figure S12.

Differing from the previous studies,[31, 38-41] the special case of $InP_3$ suggests that the interlayer interaction could be the combination of different types of interaction, and all the involved interactions need to be taken into consideration to account for the band structure evolution.

*Strength of interlayer interaction.* To quantify the overall strength of interlayer interaction in the above three materials, the interlayer interaction energy of different 2D materials, defined as the energy required to split the bilayer into two isolated unrelaxed single-layers, is calculated. The energies per unit area instead of per formula unit are calculated since various 2D materials have different lattice constants and hence different areas per cell. Thus, the interlayer interaction energy per unit area, $E_{\text{int}}$, is defined as

$$E_{\text{int}} = \frac{E_{\text{bilayer}} - 2E_{\text{single-layer}}}{A},$$

where $E_{\text{bilayer}}$ and $E_{\text{single-layer}}$ is the total energies of bilayer and unrelaxed single-layer system, respectively. $A$ is the area of the primitive cell of bilayer structure. The result of interlayer interaction energies is summarized in Table 1.

Table 1. The interlayer interaction energies (in meV/Å$^2$) of some 2D layered materials.

| Homo-occupancy interaction | Arsenene | MoS$_2$ | BP | InSe(γ) | InSe(β) | PtS$_2$ | |
|---|---|---|---|---|---|---|---|
| | 20.97 | 27.21 | 23.21 | 17.26 | 17.04 | 31.84 | |
| Hetero-occupancy interaction | SnP$_3$ | InP$_3$ | GeP$_3$ | AlP$_3$ | GaP$_3$ | CuBiS$_2$ | CuSbS$_2$ |
| | 102.58 | 87.23 | 98.01 | 135.79 | 93.34 | 67.57 | 57.09 |

From Table 1, one can find that the overall interlayer interaction energies of hetero-occupancy interaction are much stronger than those in homo-occupancy interaction. This result is consistent with the above analysis of classification. For example, the overall interlayer interaction for homo-occupancy interaction (such as that in arsenene layers) is weak because the quasi-bonding interaction destabilizes the system and competes with the vdW attraction. In contrast, for the hetero-occupancy interaction, the quasi-bonding interaction stabilizes the system and strengthen the overall interlayer interaction. We also study the interlayer interaction energies of some other 2D materials. Additionally, a recent high-throughput investigation of the exfoliation of 2D materials[60] is presented and the several representative layered materials defined as easily exfoliated in that work can all classified as homo-occupancy interaction. These results are summarized in Table 1 and suggest that: 1) our classification of interlayer interaction can give a rough estimation of the relative interlayer interaction strength, and more importantly, 2) this classification is applicable for generic 2D materials.

In summary, we classify the interlayer interaction into two main categories.

Category I: The homo-occupancy interaction. This category of interlayer interaction is relatively weak, and, it is quite common in 2D materials, such as arsenene, $MoS_2$, black phosphorus etc. In some cases, it leads to an energy gap close if the shift of bands is significant.

Category II: The hetero-occupancy interaction. This category of interlayer interaction usually does not cause a metal-semiconductor change. For example, in some cases, the involved bands are CBM and VBM, it can enlarge the band gap, which contrasts with the widely believed trend that the bandgap decreases as the layer-number increases.[41] This type of interlayer interaction is relatively strong. As special cases, we also discuss the interlayer interaction include unsaturated orbitals, which could possibly cause a metal-semiconductor transition. Considering that the structure with unsaturated orbitals in the out-of-plane direction might be unstable, this kind of interlayer interaction might induce a geometric change, as $SnP_3$ does. We also discuss the combination of these two types of interlayer interaction which is rarely mentioned in literatures.

Our classification is applicable to other typical 2D materials like TMDs, black phosphorus, InSe, etc., and provides a feasible way to estimate the relative strength of interlayer interactions. It gives a new avenue in screening 2D layered materials with different types of interlayer interaction. Our insights to the interlayer quasi-bonding interaction and interaction-induced band structure evolution deepen the understanding of the impact of interlayer interactions and could be helpful in the designing of vertical homo- and hetero-structures of 2D layered materials.

## ACKNOWLEDGMENTS

This work was supported by the National Natural Science Foundation of China (Grant Nos. 11904154 and 11774142), the Natural Science Foundation of Hebei Province of China (No. A2021201001), the Advanced Talents Incubation Program of the Hebei University (521000981390), the Guangdong Provincial Key Laboratory of Energy Materials for Electric Power under Grant No. 2018B030322001, and the Shenzhen Basic Research Fund under Grant No. JCYJ20180504165817769. Computer time was supported by the Center for Computational Science and Engineering of Southern University of Science and Technology and the high-performance computing center of Hebei University.

**Supporting Information Available:** Containing details of computational method, additional calculation and analysis of the representative materials and additional examples for the classification.